%% file: itwist20_paper_-_final.tex
\documentclass[a4paper, 10pt, twocolumn]{article}

\usepackage{amsfonts,amssymb,amsmath,amsthm}
\usepackage{subfigure}
\usepackage{graphicx}

\usepackage[top=1.5cm, left=1.5cm, right=1.5cm, bottom=1.5cm]{geometry}

\title{A mathematical approach to resilience}

\author{Dominique Pastor $^1$, Erwan Beurier $^1$, Andrée Ehresmann $^{2}$ and Roger Waldeck $^1$.\\
\footnotesize $^1$IMT Atlantique, Lab-STICC, UBL, CS 83818, 29238 Brest Cedex 3, France.\ $^2$Université de Picardie Jules Verne, CS 52501, 80025 Amiens Cedex 1.\
}
\date{\empty} 

\renewenvironment{abstract}{\bf\small {\em\ Abstract---}}{}

\newtheorem{Lemma-Definition}{Lemma-Definition}
\newtheorem{Proposition}{Proposition}

\newtheorem{Theorem}{Theorem}

\input{pastorcommands}

\begin{document}

\maketitle

\begin{abstract} In this paper, we evolve from sparsity, a key concept in robust statistics, to concepts and theoretical results of what we call the mathematics of resilience, at the interface between category theory, the theory of dynamical systems, statistical signal processing and biology. We first summarize a recent result on dynamical systems \cite{Beurier_2019}, before presenting the de-generacy paradigm, issued from biology \cite{Edelman1973} and mathematically formalized by \cite{EhresmannVanbremeersch2007, Ehresmann_2019} as the Multiplicity Principle (MP). We then make the connection with statistical signal processing by showing that two distinct and structurally different families of tests satisfy the MP.
\end{abstract}

\section{Introduction}
\label{sec:introduction}

Sparsity and robust statistics go hand-in-hand since Donoho's seminal paper \cite{Donoho94}. To a certain extent, sparsity in signal processing can be regarded as an instance of Occam's principle of parsimony (Occam's razor) because sparsity-based theories make ``{\em{the fewest unwarranted, or ad hoc, assumptions about the data from which they are derived''}} \cite{Glock2004}. Statistical decision and binary classification can also benefit from sparsity in a weaker form \cite{APM} \cite{Pastor_birkhauser} to make the connection between estimation by wavelet shrinkage and statistical decision.

Sparsity provides theories based on the least possible number of hypotheses to provide algorithms capable of dealing with very large classes of signals with guaranteed performance and, if possible, optimality. In this vein, to what extent can we discard still more hypotheses to come up with algorithms capable of performing in a complete agnostic way, that is, without any prior knowledge on the signals encapsulated in the observations made by the sensors? Otherwise said, how far can we get if we eliminate even the few hypotheses made in sparsity-based theories about the coefficients representing a certain signal, while still guaranteeing performance and even optimality?

This question is addressed in statistical testing \cite{RDT}, \cite{RDT-SUBSPACE}, \cite{RDTlm} and, more recently, in sequential statistical testing \cite{khanduri_seqRDT}, \cite{khanduri_TseqRDT}, via the notion of Random Distortion Testing (RDT). But, in the present paper, let us challenge the question itself, to open new prospects. 

Sparsity has paved a very rich way to robustness; in addition, in papers mentioned above, the decision is not only robust but even optimal. But are robustness and optimality so desirable, after all? For instance, most biological systems are resilient in the sense that, even affected by external events, (i) they can still fulfill their function, even in a degraded mode, with some performance loss, but without collapsing; (ii) they can recover their initial performance level when nominal conditions are satisfied again; (iii)  can perform corrections and auto-adaption so as to maintain essential tasks for their survival. For example, if you catch a cold, your abilities may be reduced, but in most cases, you can keep on living more or less normally and will recover. It thus turns out that, via interactions between cells, that are very simply reactive organisms, the immune system is capable of detecting, vanquishing and memorizing threats, even yet unknown. 

Can we mathematically address all these questions? In particular, can we formalize the notion of resilience, which  remains elusive, mathematically speaking, and connect it to that of robustness, which has a long history and track record in mathematical statistics \cite{Hampel86}? Many questions indeed, but already some answers that can be summarized as follows.

\section{Memoryless systems generate all discrete dynamical systems \cite{Beurier_2019}}
	
Automata model systems with inputs, internal states that can be modified, and outputs. The set of all possible states of the automaton can be regarded as the set of all potential memories. The theory of dynamical systems (DS) is particularly relevant to model automata whose state space is not necessarily finite or even countable. The state space of such a dynamical system acts as a memory of the inputs and each input alters the automaton current state. We can thus consider two extremal types of DS. First, a DS whose state space is the history of all its past inputs, appended with any new input. On the other hand, we can consider DS that forget previous inputs. These latter systems are referred to as "simple-reflex" in \cite[p. 49]{RussellNorvig2010} and memoryless in \cite{Beurier_2019}. Such a DS outputs according to its current perception of the world without storing any memory. In \cite{Beurier_2019}, it is shown that any DS system with memory is equivalent, in a strict mathematical sense, to memoryless DS suitably wired together. Such a result supports artificial neural network approaches where memory is supported by connections, and not only by the components only. The proof is constructive and underscores the crucial role played by feedback. The paper thus proves and generalizes intuitive knowledge of electronicians and computer scientists to any kind of automaton taking inputs in discrete time, for any kind of inputs and any type of outputs. 

\section{De-generacy and the Multiplicity Principle (MP)}

The results in \cite{Beurier_2019} are not limited to a very specific class of DS and the question is thus whether there exist some preferable types of memoryless DS to be used for generating DS with memory. Otherwise said, can we come up with criteria to guide our choice for memoryless DS? The answer naturally relates to our search for a formalization of the notion of resilience. To this end, biology and mathematics teach us a lot. 

On the one hand, in \cite{Edelman1973}, the authors pinpointed that ``degeneracy'' is a crucial property of living organisms. ``Degeneracy'' refers to the ability of two structurally different biological subsystems or organs to achieve functional redundancy. In the sequel, we will write ``de-generacy" instead, to emphasize the possibility to generate the same function through two structurally different subsystems, in contrast to material redundancy usually praised in engineering. De-generacy is encountered at multiple levels in living beings. For example, amino acids are the building blocks of essential proteins and are produced from “messages” included in portions of DNA. Each of these messages is called a “codon” built from three molecules, known as nucleotides. However, there are 4 possible nucleotides, which yields 64 possible codons, for 22 amino acids only. We conclude that some codons, which are structurally different by definition, provide the same amino acid: a perfect example of degeneracy. We observe that resilience seems to oppose to sparsity as it seems to require redundancy.

It then turns out that Edelman and Gally's de-generacy principle has its mathematical counterpart, namely the Multiplicity Principle (MP), introduced in \cite{EhresmannVanbremeersch2007, Ehresmann_2019} and deriving from Category Theory \cite{Spivak2014}. Through the MP, we have a complete theoretical framework to model the de-generacy principle. However, the general statement of the MP is intricate and requires some significant experience in Category Theory. Fortunately, the particular case of preordered sets suffice to derive results in statistical signal processing. More precisely, in the sequel, we will merely need the following result.

\begin{Proposition}[Multiplicity principle in a preorder]
	\label{proposition: multiplicity principle in a preorder}
	Let $\left(E, \ord \right)$ be a preorder. If there are two disjoint subsets $A,B \subset E$ such that the following conditions hold, then $E$ verifies the multiplicity principle: 
	
	\vspace{-2mm}
	
	\begin{enumerate}
		\item[(i)] $A$ and $B$ have the same sets of upper bounds
		
		\item[(ii)] There is an element $a$ of $A$ with no upper bounds in $B$
		
		\item[(iii)] There is an element $b$ of $B$ with no upper bounds in $A$
	\end{enumerate}

\end{Proposition}

%


\section{The MP in statistical signal detection}

Suppose that we observe the random vector defined on a probability space $(\Omega, \tribu, \Pbb)$ by $\VecObs{\constant}{n} = \nupleObs{\constant}{n}$
of which each component is defined on $\Omega$ and writes:
\vspace{-0.25cm}
\begin{equation}
\nonumber
\ElementaryObs_k^{[\constant]}  = \varepsilon + \Interf_k^{[\constant]}  + \ElementaryNoise_k (k \in \{1, \ldots, n\}) 
\vspace{-0.25cm}
\end{equation}
where $\varepsilon \in \{0,1\}$ stands for a binary signal, $\Interf_k^{[\constant]}$ is a random variable over $[-\constant, \constant]$ $(0 \leqslant \constant < 1/2)$ modeling some unknown interference with the signal and $\ElementaryNoise_1, \ldots, \ElementaryNoise_n \stackrel{\text{iid}}{\thicksim} \Ncal(0,1)$. At this stage, a more intricate model would not bring much and could even be detrimental to convey the main concepts and ideas underlying the approach. 

We want to estimate $\varepsilon$, that is detect whether this signal is present ($\varepsilon = 1$) or absent ($\varepsilon = 0$). We therefore want to design a test, that is, a (measurable) function $\test: \Rset^n \to \{0,1\}$ that returns an estimate of $\varepsilon$. In absence of interference ($\constant = 0$) and for a given $\level \in (0,1)$, the Neyman-Pearson (NP) test, denoted by $\Topt{n}$, is optimal because 1) it maintains the probability of false alarm $\PFA{f(\VecObs{0}{n})}$ below $\level$ and 2) it yields the largest possible probability of detection $\PDET{f(\VecObs{0}{n})}$ among all tests $f$ such that $\PFA{f(\VecObs{0}{n})} \leqslant \level$. 

In \cite{RDT} and \cite{RDTlm} are introduced an optimal test $\TRDT{n}{\tol}$ for detecting random deviations of a signal with respect to a known deterministic model in presence of independent standard gaussian noise. These random deviations can also be regarded as random distortions of the model. We thus say $\TRDT{n}{\tol}$ solves a random distortion testing (RDT) problem. The RDT problem is rotationally invariant and $\TRDT{n}{\tol}$ is optimal with respect to a specific criterion defined through suitable notions of size and power that take this invariance into account, regardless of the interference probability distribution. In particular, the size of $\TRDT{n}{\tol}$ is $\level$ for the RDT problem. The RDT test $\TRDT{n}{\tol}$ is also parameterized by $\tau$, which is chosen by the user and specifies a tolerance for distortions of the model: distortions with amplitudes below $\tau$ are asked not to be detected by $\TRDT{n}{\tol}$. 
Although not dedicated to signal detection, it can however be used to estimate $\varepsilon$ in presence of interference when $\rho \leqslant \tau < 1 - \rho$, in which case $\PFA{\TRDT{n}{\tol}} \leqslant \alpha$ to detect the presence of $\varepsilon$. 

The tests $\Topt{n}$ and $\TRDT{n}{\tol}$ can be compared as follows. First define the selectivity of a test by setting
\begin{equation}
\label{Eq: Selectivity}
\nonumber
\Selectivity{\test} \mydef \Big \{ \, \constant \in \constantinterval: \forall \VecObs{\constant}{n}, \PFA{\test ( \VecObs{\constant}{n} )} \leqslant \level \, \Big \}
\end{equation}
The selectivity shows how tolerant a test is to deviation. Let $\SetOfOracles$ be the set of oracles with level $\level$, that is the set of all $D: \{0,1\} \times \Omega \to \{0,1\}$ such that $\PFA{D} = \Pbb \left [ D(0,\cdot) = 1 \right ] \leqslant \level$ and $\PFA{D} = \Pbb \left [ D(1,\cdot) = 1 \right ] =1$. Denote by $\Mcal(\Rset^n,\{0,1\})$ the set of all tests. We define the preorder $\triangleleft$ by setting:
\begin{align*}
\forall (f,g) & \in \Mcal(\Rset^n,\{0,1\}) \times \Mcal(\Rset^n,\{0,1\}), \testa \triangleleft \testb  \, \text{iff}\\ 
	& \begin{cases}
		\Selectivity{\testa} = \Selectivity{\testb} \, \text{ and } \forall \constant \in \Selectivity{\testa}, \forall \VecObs{\constant}{n},\\
		\hspace{0.75cm} 
		\PDET{\testa \left ( \VecObs{\constant}{n} \right )} \leqslant \PDET{\testb \left ( \VecObs{\constant}{n} \right )}
	\end{cases}
	\\
\forall (f,D) & \in \Mcal(\Rset^n,\{0,1\}) \times \SetOfOracles, \testa \triangleleft D
\end{align*}
We can then prove the following statements, the third one being a consequence of the former two and Proposition \ref{proposition: multiplicity principle in a preorder}.

\begin{Theorem}[\cite{Pastor_ACT2019}]
	With the same notation as above, \medskip \\
	(i) $\Topt{n}$ and $\TRDT{n}{\tol}$ are not comparable through $\triangleleft$ \medskip \\
	(ii) $\sup_n \left( \Topt{n} \right) = \sup_n \left( \TRDT{n}{\tol} \right) = \SetOfOracles$, in the sense of $\triangleleft$ \medskip \\
	(iii) The NP tests and the RDT tests form two families of tests satisfying the MP.
\end{Theorem}

\section{Conclusions and Perspectives}
\label{Section: Conclusions}

At the interface between category theory, biology and statistical signal processing, the NP and RDT tests applied to a detection problem satisfy the MP. This opens prospects in the modeling and study of resilient systems, especially for detection through networks of sensors that could be asked to achieve the MP to guarantee resilience of the overall system in presence of interferences. It is worth noticing that the MP is achieved thanks to the RDT tests, which are not optimal for the detection task under consideration.  Our work in-progress involves extending the results above to more complex detection problems. In this respect, sequential testing problems are very appealing. On the one hand, the Sequential Probability Ratio Test (SPRT) introduced  in \cite{Wald1945} is known to be optimal for sequential testing; on the other hand, in \cite{khanduri_seqRDT}, non-optimal tests are constructed with still performance guarantees in presence of interferences. In the same way as NP and RDT tests satisfy MP, we conjecture that SPRT and the sequential tests in \cite{khanduri_seqRDT} satisfy MP as well. Another question we have in mind is the following. Between the absence of interference required for the NP test to be optimal and the RDT test that performs regardless of the interference probability distribution, can sparsity represent a trade-off making it possible to inject some prior knowledge on the interference, possibly thanks to a suitable sparse transformation?  

%
%


\bibliographystyle{abbrv}
\bibliography{Refs_v2}


\end{document}

%% file: pastorcommands.tex
\def\vec{\boldsymbol}

\def\Pbb{{\mathbb P}}

\def\Rset{\mathbb{R}}

\def\Rset{\mathbb{R}}

\def\tol{\tau}

\def\level{\gamma}

\def\Mcal{\mathcal M}
\def\Ncal{\mathcal N}

\def\test{f}
\def\testa{f}
\def\testb{g}

\newcommand{\Bcal}{\mathcal B}

\newcommand{\PFA}[1]{\Pbb_{\textsc{fa}} [ {#1} ]}
\newcommand{\PDET}[1]{\Pbb_{\textsc{det}} [  {#1} ] }

\newcommand{\ElementaryObs}{Y}

\newcommand{\VecObs}[2]{\vec{\ElementaryObs}^{[#1]}_{#2} }

\newcommand{\nupleObs}[2]{ \left( \IdealObs{#1}{1},\IdealObs{#1}{2}, \ldots, \IdealObs{#1}{#2} \right) }
\newcommand{\ord}{\preceq}

\newcommand{\Dfrak}{\mathfrak D}

\newcommand{\Topt}[1]{\test^{\textsc{np}(\level)}_{#1}}
\newcommand{\TRDT}[2]{\test^{\textsc {rdt}(\level)}_{#1,#2}}

\newcommand{\ElementaryNoise}{X}

\newcommand{\tribu}{\Bcal}

\newcommand{\reduc}[1]{\Xi}

\newcommand{\SetOfOracles}{\SymbolSetOfDecisions}
\newcommand{\mydef}{\, \, := \, \, }
\newcommand{\constant}{q}
\newcommand{\Interf}{\Delta}

\newcommand{\IdealObs}[2]{Y^{[#1]}_{#2}}

\newcommand{\SymbolSetOfDecisions}{\Dfrak}

\newcommand{\constantinterval}{[0,1/2)}

\newcommand{\Selectivity}[1]{\textup{Sel}_\level \left ( #1 \right )}

\input{./defs.tex}



%% file: defs.tex
\def\tol{\tau}

\def\Bcal{\mathcal{B}}

\def\Ncal{\mathcal{N}}

\def\Pbb{\mathbb{P}}
\def\Rset{\mathbb{R}}

\def\level{\gamma}

